\documentclass[useAMS,usenatbib,10pt]{mn2e} 
\usepackage{graphicx}
\usepackage{txfonts}

\def\HI{H\,{\sc i}}
\def\smh{$M_{\rm H_{I}}/D_{l}^{2}$}
\def\dhbdt{$D_{HI}/D_{25}$}
\def\mh1{$M_{\rm H_{I}}$}

\def\lsun{L$_{\odot}$\/\ }
\def\msun{M$_{\odot}$\/\ }

\def\kms{km s$^{-1}$}

\def\etal{{\it et al.}\thinspace}

\def\ctab#1#2#3#4{
\begin{table}
 \begin{minipage}{#1 mm}
 \label{#2}
 \caption{#3}
 {#4}
\end{minipage}
\end{table}
}

\def\cfig#1#2#3#4#5#6#7{
    \begin{figure}
    \hspace{#7cm}
    \centerline{\rotatebox{#6}{\includegraphics*[height=#3in]{#1}} }
    \vspace*{#5in}
    \caption{#2}
    \label{#4}
    \end{figure}
}

\title[\HI ~ imaging of galaxies in X-ray bright groups]{\HI ~ imaging of galaxies in X-ray bright groups }
\author [Sengupta \etal]{Chandreyee Sengupta$^{1}$\thanks{e-mail:csg@rri.res.in}, Ramesh Balasubramanyam$^{1}$\thanks{e-mail:ramesh@rri.res.in}, K. S. Dwarakanath$^{1}$\thanks{e-mail:dwaraka@rri.res.in},\thanks{Send offprint requests to: Chandreyee Sengupta} \\
$^{1}$Raman Research Institute, Bangalore 560 080 INDIA}

\begin{document}
\date{}
\pagerange{\pageref{firstpage}--\pageref{lastpage}} \pubyear{}

\maketitle

\label{firstpage}

\begin{abstract}
\noindent Environment plays an important role in the evolution of the gas contents of galaxies. Gas deficiency of cluster spirals and the role of the hot intracluster medium (ICM) in stripping gas from these galaxies is a well studied subject. Loose groups with diffuse X-ray emmision from the intragroup medium (IGM) offer an intermediate environment between clusters and groups without a hot IGM. These X-ray bright groups have smaller velocity dispersion and lower temperature than clusters, but higher IGM density than loose groups without diffuse X-ray emission. A single dish comparative study of loose groups with and without diffuse X-ray emission from the IGM, showed that the galaxies in X-ray bright groups have lost more gas on average than the galaxies in non X-ray bright groups. In this paper we present GMRT \HI~ observations of 13 galaxies from 4 X-ray bright groups: NGC5044, NGC720, NGC1550 and IC1459. The aim of this work is to study the morphology of \HI~ in these galaxies and to see if the hot IGM has in any way affected their \HI~ content or distribution. In addition to disturbed \HI~morphology, we find that most galaxies have shrunken \HI~disks compared to the field spirals. This indicates that IGM assisted stripping processes like ram pressure may have stripped gas from the outer edges of the galaxies.  

\end{abstract}


\newpage
\section {Introduction}\noindent Galaxies are found in different environments like field, loose groups, compact groups and clusters. The environment plays an important role in the gas contents of galaxies. A range of gas removal processes work in these environments. Depending on the position of the galaxy in the group or in the cluster, the galaxy looses gas to the environment to different degrees. Spiral galaxies in cluster cores are well known to be atomic hydrogen (\HI)~deficient. They move through the dense hot cluster cores with high velocities and lose gas through ram pressure stripping \citep{RamP}. The hot ($10^{8}$ K) intracluster medium (ICM) can also strip gas from the galaxy through evaporation \citep{Evap}. Detailed observations have been carried out over the past three decades to probe gas deficiency in cluster galaxies. They seem to suggest that a large fraction of the cluster galaxies are actually deficient in \HI~.  More deficient galaxies are found closer to the cluster centre and many galaxies have truncated or shrunken \HI~disks or extraplanar gas indicating ram pressure stripping (\cite{H1Def}, \cite{cayatte1990}, \cite{crowl}). 

\vspace*{0.1cm}

\noindent While cluster galaxies have been extensively studied, not many studies exist about the gas content and gas deficiency of the galaxies in groups and the possible gas removal processes that can succesfully work in such environments. Compact groups have been studied better than loose groups. A sample of 72 HCGs were studied with single dish and a subset of those with the VLA. The galaxies in the HCGs were found to be \HI~deficienct by greater than a factor of 2 on an average. Some of these groups have been detected in X-rays and the X-ray detection rate is found to be higher for the \HI~ deficient groups \citep{H1HC1}.

\vspace*{0.1cm}

 \HI~content and deficiency studies of loose groups are not so well reported in literature. \HI~deficiency by a factor of more than 1.6 has been reported in some members of a loose group in the Puppis region \citep{H1NC1}. Galaxies in Eridanus, a large loose group, have been found to be \HI~deficient by a factor of 2 to 3. About 31 galaxies from the Eridanus group have been imaged in \HI~ using the Giant Meterwave Radio Telescope (GMRT). From these images and the overall group properties, tidal interaction was found to be the most likely gas removal mechanism responsible for the \HI~deficiency of Eridanus galaxies \citep{eridan1}.

\vspace*{0.1cm}

\noindent 
Ram pressure becomes an effective gas removal process only when the \HI ~surface mass density is less than $\rho_{0}$v$^{2}/(2\pi~G\sigma_{*})$, where $\sigma_{*}$ is the stellar surface mass density. So, larger the intra-cluster medium (ICM) density, $\rho_{0}$, and galaxy velocity, v, the more effective is this stripping. Thus in the absence of a hot IGM and a high velocity dispersion, tidal interaction appears to be the most favourable process by which galaxies in groups can lose gas, though there are some cases where ram pressure is also acting in groups \citep{kantharia}. Detection of hot intragroup medium (IGM) in some poor groups has presented an intermediate environment between clusters and groups without a hot IGM. When the group has a hot IGM, ram pressure or thermal conduction can contribute significantly to make the galaxies deficient in \HI. Since low velocity dispersion in these groups increases the probability of tidal interactions, tidal assisted ram pressure stripping can also contribute to gas loss from galaxies \citep{TRamP}. NGC2276, a spiral in the NGC2300 group is found to be \HI~ deficient and signatures of the hot IGM affecting the galaxy were detected \citep{rasmussen}. A comparative single dish study of the \HI~contents of loose groups with and without the hot IGM indicates that the X-ray emitting medium may play a significant role in removing gas from the galaxies, as galaxies in these groups are found to be more \HI~deficient than the galaxies in groups without X-rays \citep{paper1}. Thus it is interesting to study the \HI~contents and morphologies of galaxies in groups with a hot IGM, to see if IGM assisted stripping has left its signatures in the \HI~distribution of the galaxies. 
 
\vspace*{0.1cm}

\noindent In this paper we present GMRT \HI~ images of nine spiral galaxies in four groups whose IGM is detected in X-rays. The aim of this work is to study the morphology of \HI~ in these galaxies and to see if the hot IGM has in any way affected their \HI~ content and distribution.

\section {GMRT observations and Data analysis}

\noindent Thirteen galaxies from four groups, where the IGM has been detected in X-rays, were observed with GMRT in the \HI~ 21cm line. The four groups from which this sample of thirteen galaxies were drawn are NGC5044, NGC1550, NGC720 and IC1459. They have a varied range of X-ray properties (Table 1). NGC5044 is a well studied bright X-ray group with the diffuse emission being mostly circular in morphology with concentric isophotes \citep{chandra1}. NGC1550 is amongst the brightest X-ray groups so far obsevred and has more relaxed X-ray morphology, except some assymetries near the group centre, than the other $>$ 1keV groups like NGC5044 \citep{1550}. NGC720 and IC1459 are moderately bright groups and are also comparatively nearer than the other two. All the groups have memberships $\sim$~10 to 20 and have at least one bright elliptical galaxy at their X-ray centres. GMRT \HI~observations of thirteen galaxies from these four groups are presented in this paper. Table 1 lists the group names, their distances, X-ray luminosities and temperatures of the IGM and the number of galaxies observed with the GMRT. 

\vspace*{0.1cm}

\noindent GMRT is an interferometric array of 30 antennas, each of 45m diameter, spread over a maximum baseline of 25 km. At frequencies of $\sim$~1400 MHz, the system temperature and the gain (K/Jy) of the instrument are 76K and 0.22 respectively. The full width at half maximum of the primary beam of GMRT antennas in L band (frequency 1.4 MHz) is 24 \arcmin. The FX correlator of GMRT offers a total of 128 channels for spectral line observations in the full polar mode. The sources selected have a wide range in \HI~ content. The motivation was to study if there were any evidance of the hot IGM being responsible for gas loss from these galaxies. All the sources are below 5\arcmin ~in optical size, their distances are between 20 to 60 Mpc, and their declinations range from -36 $^{\circ}$ to +02 $^{\circ}$. With GMRT they were studied at a spatial resolution of $\sim$3 kpc, a velocity resolution of $\sim$13 km/s and to a \HI~ column density of $\sim$ 1$\times 10^{20}$ per cm$^{2}$. Table 2 lists the relevant details of the observations carried out with the GMRT. 

\noindent Data obtained with the GMRT were reduced using AIPS (Astronomical Image Processing System). The procedure used to get the total \HI~ maps and the velocity fields is the following. Bad data (dead antennas, antennas with significantly lower gain than others, RFI) were flagged and the data was calibrated for amplitude and phase using standard primary and secondary calibrators. The spectral responses of the filters were corrected by calibrating the data using a standard primary calibrator or when availbale a strong secondary calibrator. The calibrated data were used to make both the \HI ~line images and the 20 cm radio continuum images. The continuum data set were prepared from the calibrated data using the line (\HI~ emission) free channels (frequencies). The data were then averaged in frequency and self calibrated. AIPS task 'IMAGR' was then used to get the deconvolved high resolution contunuum images. For the \HI~ line images the calibrated data were continuum subtracted using AIPS tasks 'UVSUB' (used to subtract a model, in this case the clean components from the continuum image, from the uv data) and 'UVLIN' (used to subtract continuum by making linear fits to the uv data). The task 'IMAGR' was then used to get the final 3 dimensional deconvolved \HI~data cubes. From these cubes the total \HI~ images and the \HI~ velocity fields were extracted using the AIPS task 'MOMNT'. By definition the total \HI~maps are the integrated flux density maps,


\begin{equation}
I_{tot}(\alpha,\delta)=\triangle \rm{v}~ \sum_{i=1}^{Nchan}S_{\nu}(\alpha,\delta,\nu_{i})
\end{equation} 

\noindent and the velocity fields are the intensity-weighted first order moment of the \HI~distribution at different velocities.


\begin{equation}
\rm{\overline v}= \frac{\sum_{i=1}^{Nchan} v_{i}~ S_{\nu}(\alpha,\delta,\nu_{i})}{\sum_{i=1}^{Nchan}S_{\nu}(\alpha,\delta,\nu_{i})}
\end{equation}


\noindent Here $S_{\nu}(\alpha,\delta,\nu_{i})$ is the observed flux density at the position ($\alpha$ ,$\delta$) in channel i (which has frequency $\nu_{i}$ and velocity v$_{i}$, $\triangle$v is the velocity resolution of the channels and is a constant for a particular dataset and Nchan is the total number of channels over which the \HI~ line exists. The total \HI~maps were produced at an angular resolution of 15$\arcsec$ to 25$\arcsec$ depending upon the distance to the galaxies, so that the \HI~morphology could be studied at $\sim$3 to 4 kpc resolution. Some of the velocity fields were produced at similar resolutions (15$\arcsec$ to 25$\arcsec$) and some were at lower resolutions (listed in table 2.). The total \HI~ and the velocity fields overlaid on the respective optical images of all the \HI~ detected galaxies are presented in this paper. The rms per channel of the line cubes and the synthesised beam sizes are listed in table 2. The units of total \HI~ maps have been converted from Jy m/s (the original map unit) to \HI ~column densities using the formula,


\begin{equation}
N(HI)={\frac{1.1\times10^{21}~cm^{-2}}{\theta_{a} \times \theta_{b}}}\triangle \rm{v} \sum_{i=1}^{Nchan}S_{\nu}(\alpha,\delta,\nu_{i})
\end{equation}  


\noindent where $\theta_{a}$ and $\theta_{b}$ are the major and minor axes of the synthesised beam in arcseconds, $\triangle$v is the velocity resolution of the channels in km/s and $S_\nu(\alpha,\delta,\nu$) is the \HI~flux density at the position ($\alpha,\delta$) in the channel $i$ in mJy.

\section {Results}
\subsection {\HI~content and deficiency}

\noindent  The total \HI~ maps of the nine \HI~ detected galaxies and their respective velocity fields are presented in Fig 1 - 20. The figure captions contain the galaxy names and the \HI~column density levels plotted (in total \HI~maps) or the \HI ~velocity contours (in the \HI~ velocity field maps). Following are a set of short notes on each of these nine galaxies. 

\vspace*{0.2cm}

\noindent{\bf{\it{ARP004 (Fig 1, 2, 3)}}}: An irregular galaxy from the NGC720 group. The \HI~distribution is normal in low resolution (Fig 1) and extremely fragmented in high resolution (Fig 3). But the galaxy is not \HI~deficient. Velocity field is apparently normal. This galaxy is detected in FIR and not detected in radio 20 cm and has an FIR excess.

\vspace*{0.2cm}

\noindent{\bf{\it{DDO015 (Fig 4, 5)}}}: A dwarf galaxy from NGC720 group. An extremely \HI~deficient galaxy. Most part of the western and central regions have depleted \HI~content. The \HI~disk is also truncated.

\vspace*{0.2cm}
 
\noindent{\bf{\it{MCG-03-34-04 (Fig 6, 7) }}}: This galaxy belongs to the group NGC5044. Its morphological classification is S0 (+:), where the '+' sign indicates a late type lenticular and a ':' sign indicates uncertain morphological assignment. Normally lenticulars do not contain much \HI. But this galaxy has a huge \HI~mass $\sim$ 8.0$\times~ 10^{9}$ \msun. A galaxy of this size and morphological type S0, should contain $\sim$ 1.8$\times~ 10^{9}$ \msun. This galaxy is detected in 20 cm radio continuum and FIR. About 5.5$\arcmin$ away from this galaxy is NGC4997, which is non-detected in both \HI~and radio continuum, and 11.8$\arcmin$ away is 2MASX J13085477-1636106, a galaxy which has been detected in \HI~ in this GMRT observation. But since 2MASX J13085477-1636106 is close to the edge of the primary beam, no further study can be done with the present observations. Since S0 galaxies have an uncertainty about their \HI~content and many galaxies even do not contain any \HI, though an \HI~deficiency has been quoted for MCG-03-34-04 (table 4), no gas loss by ram pressure or evaporation has been calculated for this galaxy. 

\vspace*{0.2cm}

\noindent{\bf{\it{MCG-03-34-41 (Fig 8, 9)}}}: A late type spiral from  NGC5044 group. The HI~disk is truncated compared to normal spirals and the galaxy has moderate \HI~deficiency. It is FIR detected and radio non-detected and has an FIR excess.
\vspace*{0.2cm}

\noindent{\bf{\it{SGC1317.2-1702 (Fig 10, 11)}}}: A late type spiral in NGC5044 group. The \HI~distribution is clumpy. The galaxy is moderately \HI~deficient with a shrunken \HI~disk.  

\vspace*{0.2cm}

\noindent{\bf{\it{SGC 1316.2-1722 (Fig 12, 13)}}}: A late type spiral in NGC5044 group. The is an \HI~deficient galaxy with a shrunken \HI~disk. About 6.1$\arcmin$ away towards north-west, \HI~ has been detected in another galaxy KDG 220. This is a dwarf galaxy. \HI ~has been detected in this galaxy by GMRT at a velocity of 2470 km/s. Radial velocity of this galaxy was unknown before this observation. Total \HI~map of SGC 1316.2-1722 also shows KDG 220, in the north-west corner.  

\vspace*{0.2cm}

\noindent{\bf{\it{UGC3014 (Fig 14, 15)}}}: This spiral belongs to NGC1550 group. It is close to NGC1550 in both velocity and coordinate, but not a group member according to \cite{LGGC}. \HI~ distribution has an extension in the eastern side and a shrunken \HI~disk. Velocity field of this galaxy is normal. This galaxy is detected in radio continuum and FIR.

\vspace*{0.2cm}

\noindent{\bf{\it{UGC3004 and UGC3005 (Fig 16, 17, 18) }}}: These two galaxies belong to NGC1550 group. UGC3004 (V$_{\rm {opt}}$=3571 km/s) is recognised as a group member, but UGC3005 (V$_{\rm {opt}}$=3215 km/s) and UGC3006 (V$_{\rm {opt}}$=3664 km/s) are not listed group members \citep{LGGC}. UGC3006 was neither detected in \HI~ nor in radio continuum. UGC3004 and UGC3005 look like interacting galaxies in their total \HI ~maps, but they are actually well seperated in velocity and are probably not interacting. UGC3004 has been previously observed by the VLA in \HI~and in 6 cm \citep{chaboyer}. The spectral index of this galaxy was found to be -1.6. This galaxy is also IRAS detected. UGC3005 is a low surface brightness edge on spiral with normal \HI~content.

\vspace*{0.2cm}

\noindent{\bf{\it{IC5269B (Fig 19, 20)}}} : A barred spiral galaxy from IC1459 group. \HI~distribution and velocity field is normal. 

\vspace*{0.3cm}

\noindent A possible way to establish whether a galaxy has lost gas is to compare its gas content with that of the field galaxy sample, of the same morphological type. The parameter that measures whether a galaxy has an excess or a lower \HI~content compared to a field sample, referred to as \HI~deficiency, is defined as \citep{H1Def},

  
\begin{equation}
{\it def_{\rm {H_{I}}}=log{{\frac{M_{H_{I}}}{D_{l}^{2}}}|_{field}}~-~log{{\frac{M_{H_{I}}}{D_{l}^{2}}}|_{obs}}} 
\end{equation}

\noindent where $M_{H_{I}}$ is the total \HI~mass of a galaxy and $D_{l}$ is the optical major isophotal diameter (in kpc) measured at or reduced to a surface brightness level m$_B$ = 25.0 mag/arcsec$^{2}$.
\vspace*{0.1cm}
 
\noindent The expected field values of \smh ~for various morphological types are taken from \cite{H1Def}. While \cite{H1Def} used the UGC blue major diameters for $D_{l}$, in this work RC3 major diameters have been used. 
A hubble constant of 75 \kms ~Mpc$^{-1}$ has been used throughout the paper to get distances to the galaxies from their optical velocities. To take care of the difference in the surface matter densities that result from the use of RC3 diameters, a value of 0.08 \citep{RC3UGC} has been added to the expected surface matter densities given by \cite{H1Def}. The relevant values are listed in table 3. 

{\begin{table}
\caption{Groups Observed}
\begin{tabular}{llllr}
\hline
 Group  &Distance & T$_{X}$ & L$_{X}$ &galaxies \\ 
         & (Mpc)    & (K)       & (erg/s)   & (\#)  \\
\hline
NGC720 &21.5 &0.51 &7.24$\times 10^{40}$&2  \\ 
NGC1550&48.0 &1.37 &2.0$\times10^{43}$ & 4  \\
NGC5044&34.8 &1.02 &6.45$\times 10^{42}$& 6 \\
IC1459 &23.6 &0.63 &3.31$\times 10^{40}$& 1\\
\hline
\end{tabular}
\end{table}
}

\noindent  The \HI~ mass was derived from the integrated flux density using the following formula,

\vspace*{0.1cm}

\begin{equation}
$\mh1(\msun)$ = 2.36 \times 10^{5} \rm{D^{2}} \triangle\rm{v} \sum_{i=1}^{Nchan}S_{i}
\end{equation}
\vspace*{0.1cm}
\noindent where D is the distance to the galaxy in Mpc and $\triangle$v is the velocity resolution in km/s.
\noindent The integrated flux densities for this purpose were taken from the \HI ~Parkes all sky survey (HIPASS) catalog (if not available in catalog, from the spectra) and for galaxies which had noisy HIPASS spectra, GMRT spectra were used to calculate their \HI~masses. The \HI~masses and \HI~deficiencies for the sample galaxies estimated in the above mentioned way are listed in table 4.


\subsection {\HI~diameter to optical diameter ratio}
 
\noindent Due to large uncertainties of the \HI ~surface density values of the field galaxies (please see table 3), it becomes difficult in certain cases to judge if a galaxy is \HI~ deficient compared to the field sample. An alternate way of assessing whether a galaxy has lost gas is by comparing its $D_{\HI}/D_{25}$ ratio to the average value of this ratio found in spirals and irregulars. $D_{\HI}/D_{25}$ is the ratio of the \HI~ diameters measured at face-on \HI~surface density of 1\msun per $pc^{2}$ to the optical major isophotal diameter measured at or reduced to a surface brightness level m$_B$ = 25.0 mag/arcsec$^{2}$. The $D_{25}$ in this section is the same as the $D_{l}$ of section 3.1. The high resolution (15$\arcsec$ to 20$\arcsec$) total \HI~ maps of the galaxies obtained using AIPS were deprojected using the task 'ELLINT' of GIPSY (Groningen Image Processing System). The resultant \HI~ surface density profiles were then fitted with a gaussian \citep{chamaraux} and the \HI~ diameters were then measured at or to an extrapolated face-on \HI~ surface density of 1\msun per $pc^{2}$. The values of $D_{\HI}/D_{25}$ for all the \HI~detected galaxies are listed in table 4. The high resolution image of ARP004 did not yeild any significant surface density profile and thus the $D_{\HI}/D_{25}$ value of this galaxy is absent from table 4. They are to be compared with an average value found in spirals and irregulars in fields and groups. \cite{eridan1} and \cite{broeils} find the average $D_{\HI}/D_{25}$ value for spirals and irregulars to be 1.7$\pm$ 0.80 and 1.7$\pm$ 0.05 respectively. The average value of $D_{\HI}/D_{25}$ for the eight galaxies in this sample is 1.1$\pm$0.12. A plot of \HI~deficiency against $D_{\HI}/D_{25}$ for all the galaxies is also presented (Fig. 24). The plot shows decrease in \HI~deficiency with increasing $D_{\HI}/D_{25}$. 


\vspace*{0.1cm}

\subsection {Radio Continuum}

\noindent Three of the thirteen galaxies observed with the GMRT were detected in radio continuum. All three had previously been detected in the NRAO VLA Sky Survey (NVSS). One of the three galaxies (MCG-03-34-04) is listed as a lenticular in literature, while the other two (UGC3004 and UGC3014) are listed as spirals with no further morphological specification. High resolution 20 cm GMRT maps of these three sources are presented in this paper (Fig. 21, 22, 23). Table 5 lists the relevant parameters. 
 
\noindent Five galaxies in this sample are FIR detected. The FIR luminosities were derived from the 60${\mu m}$ flux using the following relation \citep{yun}. 

\begin{equation}
\log L_{60\mu m}(L_{\odot})= 6.014+ 2\log D + \log S_{60\mu m}
\end{equation} 

\noindent where $S_{60\mu m}$ is FIR 60${\mu m}$ flux density in Jy, $D$ is the distance to the galaxy in Mpc and $L$ is the 60${\mu m}$ FIR luminosity. For $S_{60\mu m}$, in all the five cases, IRAS (Infrared Astronomical Satellite) 60${\mu m}$ fluxes were used. The 20cm radio continuum luminosities were derived using \citep{yun}

\begin{equation}
\log L_{1.4GHz}(WHZ_{_1})=20.08 +2\log D +\log S_{1.4GHz}
\end{equation}

\noindent where $S_{1.4GHz}$ is the radio 20cm flux density in Jy, $D$ is the distance to the galaxy in Mpc and $L_{1.4GHz}$ is radio luminosity in 20 cm. For the two FIR detected but radio non-detected galaxies (ARP004 and MCG-03-34-41), 3$\sigma$ radio upperlimit fluxes were used in calculating their luminosities. The three galaxies which are detected both in radio and FIR follow the FIR-radio correlation. The radio non-detected galaxies deviate from the radio-FIR correlation to some extent. To quantify this deviation the $q$ parameter was calculated, using the following relation \citep{condon}

\begin{equation}
q=\log(\frac{FIR}{3.75\times 10^{12} W m^{-2}})- \log(\frac{S_{1.4GHz}}{1.0\times 10^{26} W m^{-2} Hz^{-1}})
\end{equation}

\noindent where $S_{1.4GHz}$ is the 20 cm radio flux density in Jy and $FIR$ is 
calculated in the following manner \citep{helou}
\begin{equation}
FIR= 1.26 \times 10^{-14} (2.58 S_{60\mu m} + S_{100 \mu m}) W m^{-2}
\end{equation} 
\noindent The plot of $q$ versus $\log L_{60\mu m}$ (Fig 25), shows the three galaxies, detected both in radio and FIR, near the $q$=2.34 line, which is the mean value of $q$ for spirals and irregulars \citep{yun}. The two dotted lines about the $q$=2.34 line represent 3 times radio excess (bottom line) and 3 times radio deficit (top line) respectively. The two galaxies, ARP004 and MCG-03-34-41, represented with arrows for radio non-detection, close to the upper dotted line, tend to show they have excess FIR emission. Star formation rates of the three radio detected galaxies were calculated using \citep{yun}  

\begin{equation}
SFR (M_{\odot} yr^{-1}) = 5.9 \pm 1.8 \times 10^{-22} L_{1.4GHz} W Hz^{-1}  
\end{equation} 
 
\noindent where $L_{1.4GHz}$ is the radio luminosity in 20 cm, detected by GMRT.
The star formation rates are normal in these galaxies ranging from $\sim$~1 to 2 \msun $yr^{-1}$. The relevant parameters of the radio continuum and FIR properties of the five galaxies are listed in table 5.



\section {Discussion}

\noindent Environment affects the gas content and morphology of a galaxy as demonstrated by studies of cluster galaxies (\cite{H1Def}, \cite{cayatte1990}). Different gas removal processes like ram pressure stripping, galaxy harassment (\cite{moore}), strangulation (\cite{PreProc2}) and thermal conduction (\cite{Evap}) can strip off gas from galaxies. Tidal interactions are less likely to work as an efficient gas removal mechanism in the cluster environment as the velocities with which the galaxies move past each other are very high. 
In groups, the environment is diffierent from the clusters. The IGM densities are often lower than the ICM densities and the velocity dispersion of the galaxies in groups are also lower than in clusters. The lower velocity dispersion helps tidal interactions to work better in groups and at the same time makes ram pressure stripping less efficient process than in clusters. An effort has been made here to estimate how much gas loss is possible through ram pressure stripping and thermal conduction, and to see whether that can explain the observed gas deficiency and shrunken \HI~ disks.

\vspace*{0.1cm}
  
\subsection{Expected HI deficiencies}

\subsubsection{Ram Pressure}

\noindent In cluster environments ram pressure stripping is seen to be an effective process for removing gas from galaxies. In groups this process was not considered to be an efficient one because of lower IGM density and lower velocity dispersion, both being smaller by an order of magnitude compared to clusters. These make direct ram pressure
stripping effective only below a critical \HI~ column density of $10^{19}$ per $cm^{2}$ for a normal galaxy with an optical radius of 10 kpc, and $10^{11}$ stars . However in X-ray bright groups this picture can be different. The dense X-ray gas in the central regions of the groups can strip off a large amount of neutral gas from the galaxies which move across these regions. A simple estimation is done here for all the \HI ~detected galaxies to study the effect of ram pressure on them. Ram pressure stripping \citep{RamP} is effective for a galaxy when the \HI ~surface mass density is less than
$\rho_{0}{\rm v}^{2}/(2\pi~G\sigma_{*})$, where $\sigma_{*}$ is the stellar surface mass density, $\rho_{0}$ is the IGM density and $v$ is the velocity with which the galaxy is moving through the medium. Stellar surface matter density for these galaxies were estimated using their K-band magnitudes and K-J colors. In the absence of K and J band data, other bands were used. The Mass to light ratio in K band ( $M/L_{K}$ ) of the galaxies are related to the K-J colors by the formula
 \citep{bell}
\vspace*{0.1cm}

\begin{equation} 
log(M/L_{K})= a_{K}+ b_{K} ~color_{K-J} 
\end{equation}
\vspace*{0.1cm}
\noindent Luminosity in K-band, $L_{K}$ is related to the absolute magnitude in K-band ($M_{K}$) in the following way \citep{worthey}

\begin{equation}
L_{K}(L_{\odot})= \exp {(0.921034~ (3.33 - M_{K}))}
\end{equation}

\noindent Thus the stellar mass surface density ($\sigma_{*}$) for all the galaxies for which the necessary photometric observations were available, were estimated using their magnitude and color information. For $\rho_{0}$, the local IGM densities at the projected positions of the galaxies were used. Since all the groups in this sample do not have published density profiles for the IGM, the local densities for those cases (NGC720 and IC1459) were taken to be an average of the densities at the galaxy positions of two well studied groups in X-rays, NGC1550 and NGC5044, which also belong to this sample (\cite{1550}, \cite{5044density} ). The velocity dispersions of the groups were used as the velocities with which the galaxies are moving through the medium (v). For NGC1550 and NGC5044 the group velocity dispersions were taken from the literature (\cite{1550} and \cite{cellone}). Such detailed study are not present for the groups NGC720 and IC1459. Therefore velocity dispersion for these two groups were derived using the IGM temperatures ($T$) found from the X-ray observations \citep{DifX}. Thus dispersion for these groups is basically $\sqrt{\frac{3kT}{m_{p}}}$, where $m_{p}$ is the proton mass and $k$ is the Boltzmann constant. Thus the critical \HI~column density $\sigma_{\mu}$, beyond which ram pressure is able to stirp off gas if a galaxy moves face on through the IGM, were calculated for these galaxies using

\vspace*{0.1cm}
 
\begin{equation}
\sigma_{\mu}= \rho_{0}{\rm v}^{2}/(2\pi~G\sigma_{*})
\end{equation}

\vspace*{0.1cm}

\noindent Assumimg the \HI~ distribution to be of constant thickness and to be distributed in a Gaussian profile \citep{chamaraux} 
\begin{equation}
\sigma(r)= \sigma_{0}~ 2^{-{r^{2}}/{r_{H}^{2}}}\end{equation}

\noindent where $r_{H}$ is the radius within which half the \HI~ mass is present, the mass outside the radius of the critical \HI~column density ($\sigma_{\mu}$), for a galaxy was calculated in the following way. The total expected \HI~mass (\mh1) of a galaxy (of a known size and a morphological type) was derived using the field values \citep{H1Def}. The $r_{H}$ of equation 14. was then found by solving

\begin{equation}
\int_{0}^{\infty} 2\pi r \sigma(r) dr = $\mh1$
\end{equation}

\noindent which simplifies to 

\begin{equation}
r_{H}= \sqrt {\frac {2\log 2 M_{\rm H_{I}} }{2\pi\sigma_{0}}} 
\end{equation}
 
\noindent The next step was to find the \HI~ radius (R) corresponding to the critical \HI~column density ($\sigma_{\mu}$). This was done by solving

\begin{equation}
2^{-{{R^{2}}/{r_{H}^{2}}}}= \sigma_{\mu}/ {\sigma_{0}}
\end{equation}

\noindent In the absence of the true peak \HI~column density ($\sigma_{0}$) values, the peak \HI~column density values of the GMRT total \HI~images were used, assuming that the central regions of the galaxies have been unaffected by the gas removing processes. Finally the \HI~mass ($M_{lost}$) outside the \HI~ radius ($R$) corresponding to $\sigma_{\mu}$ was calculated by integrating the \HI~ column density profile from $R$ to $\infty$. 
 
\begin{equation}
2\pi \sigma_{0} \int_{R}^{\infty} r 2^{-{{r^{2}}/{r_{H}^{2}}}} dr = M_{lost}
\end{equation}

\noindent The percentage mass loss by this process ($M_{lost}$/\mh1) for all the galaxies for which the relevant data were available are listed in table 4, column 6. 

\noindent However these estimates were done under the following assumptions,\\
\begin{itemize}
\item{the galaxy is moving face-on through the IGM}
\item{the position of the galaxy with respect to the group centre is same as the projected distance from the group centre}
\item{the orbital history of the galaxy is not taken into account, i.e whether the galaxy has crossed the group centre or not.}
\end{itemize} 
 These may explain why the individual \HI~deficiencies of the galaxies and their percentage ram pressure stripped gas do not necessarily match (table 4). Still these calculations show that ram pressure by itself or tidal aided ram pressure cannot be ruled out as important gas removing processes in the groups having a hot IGM.



\vspace*{0.1cm}

\subsubsection{Thermal conduction}

\noindent Evaporation via thermal conduction is another process which is responsible for mass loss from galaxies embedded in a hot medium \citep{Evap}. For classical, unsaturated thermal conduction the rate of mass loss is 

\vspace*{0.1cm}

\begin{equation}
$\.{M}$ = 700~ M_{\odot}~ yr^{-1}~ ( \frac{T_{IGM}}{10^{8}} )^{5/2}~(\frac{R}{20kpc})~(\frac{\ln \Lambda}{40})^{-1}
\end{equation}

\noindent where $T_{IGM}$ is the IGM temperature in Kelvin, R is the radius of the galaxy and $\ln \Lambda$ is the Coulomb logarithm (\cite{Evapfor}, \cite{valluri}). In these estimates, $T_{IGM}$ values are taken from the X-ray observations of the groups, R is the optical size (${\rm D_{l}}/2$) of the galaxy and Coulomb logarithm is assumed to be unity. Mass loss possible through unsaturated thermal conduction in 1 Gyr is listed in table 4. Column 8 of table 4 lists the total expected \HI~ deficiency values, a sum total of columns 6 and 7. These estimates are done under the following assumptions,

\begin{itemize}
\item{thermal conduction has been considered to be unsaturated}
\item{the galaxies are embedded in the same temperature for 1 Gyr}
\item{there is no effect of magnetic fields  in supressing the mass loss rate}

\end{itemize}
\vspace*{0.1cm}
\noindent Because of these assumptions, these estimates may not necessarily reflect the true gas loss for the galaxies through this process. In practice thermal conduction is saturated on galaxy scales, as the mean free path of the electrons in the hot plasma is comparable to the size of a galaxy. Saturated thermal conduction along with the local magnetic fields will reduce the gas loss through thermal conduction significantly. Also the assumption that the galaxies are embedded in the same temperature for 1 Gyr, may not be always correct. However, within these limitations, these calculations demonstrate that under some reasonable assumptions, ram pressure stripping and in some cases thermal conduction can actually strip off gas from galaxies in these groups.



\vspace*{0.1cm}

\subsection{Observed \HI~properties}
\noindent The galaxies were observed with the aim to study their \HI~morphology and to see if IGM assisted gas stripping has left any signature on their \HI~distribution. The \HI~deficiency estimated for each \HI~ detected galaxy (eqn.4) are listed in table 4. Most of the galaxies which are reasonably \HI~deficient, by about a factor of 2 or more, show distorted \HI~morphology. DDO015 is \HI~deficient by a factor of 9, and the total \HI~ image of this galaxy shows that gas is mostly present in the eastern region. Part of western and central region of this galaxy has depleted gas content. For SGC1317.2-1702 \HI~distribution is clumpy and fragmented. This galaxy is deficient by a factor of 3. ARP004 is not a \HI~deficient galaxy, but shows an extremely fragmented \HI~structure in the high resolution map (Fig. 3). But in a lower resolution map the galaxy looks very different with a normal \HI~distribution (Fig. 1). However this is the only case where the high and low resolution maps look very different and therefore both the maps are presented in this paper. Another interesting galaxy in this sample is MCG-03-34-04. This is classified as a lenticular (S0) in literature. Lenticulars normally do not contain much \HI~. An average field lenticular of this size is expected to contain an \HI~mass of 1.8$\times 10^{9}$ \msun whereas this galaxy has a \HI~ mass of about 8$\times 10^{9}$ \msun.

\vspace*{0.1cm}

\noindent The ratio $D_{\HI}/D_{25}$, also gives an idea of gas loss from galaxies. The average value of $D_{\HI}/D_{25}$ for spirals and irregulars in the field has been found to be 1.7 $\pm$~0.8. The average of the current sample is 1.1$\pm$0.12, with most galaxies having a value close to 1.0 (table 4). This points towards the possibility of ram pressure stripping off the low column density gas from the outer edges of the galaxies. Since these galaxies belong to groups where velocity dispersion is low enough to allow tidal interactions, tidal aided ram pressure stripping can also remove gas from the galaxies \citep{TRamP}. In this case tidal interaction can stretch the gas below the critical \HI~ column density and then even a mild ram pressure can strip off the extended low column density gas from the galaxies. All these mechanisms can give rise to shrunken or truncated \HI~ disks. MCG-03-34-41 is a spiral galaxy deficient in \HI~by a factor of 3. The \HI~morphology of this galaxy seems to be apparently normal (Fig.8) other than the fact that its $D_{\HI}/D_{25}$ value is 1.1, which is significantly less than galaxies with normal gas content. The plot of \HI~ deficiency versus $D_{\HI}/D_{25}$ (Fig.24), indicates that $D_{\HI}/D_{25}$ is smaller for galaxies with higher \HI~deficiency. 

\vspace*{0.1cm}

\section {Conclusion}
A single dish comparative study revealed that galaxies in X-ray bright groups have lost more gas on average than the galaxies in non X-ray bright groups. To study if the hot IGM is responsible for gas stripping from these galaxies, 13 galaxies from 4 X-ray bright groups were imaged in \HI ~with GMRT. Disturbed \HI~ morphology was seen in some cases but most galaxies were seen to have a shrunken \HI ~disk. This indicates that ram pressure may have stripped gas from the outer edges of the galaxies. In a group environment where tidal interactions can work better than in clusters and ram pressure alone cannot strip gas as efficiently as in clusters, tidal aided ram pressure can strip off gas from the outer edges of the galaxies, giving rise to apparently undisturbed but shrunk \HI ~disks. 

\section*{Acknowledgments}
\noindent The GMRT is operated by the National Centre for Radio Astrophysics of the Tata Institute of Fundamental Research. This research has made use of the NASA/IPAC Extragalactic Database (NED) which is operated by the Jet Propulsion Laboratory, California Institute of Technology, under contract with the National Aeronautics and Space Administration. We thank R.A. Angiras for his help in using some of the GIPSY routines.

\onecolumn

{\begin{table}
\caption{GMRT Observations }
\begin{tabular}{llllllllr}
\hline
 Galaxy&Coordinate &Optical  & M type & $\tau$ & Bandwidth &rms per channel&beam size (") & beam size (") \\ 
&(J2000)&Velocity (km/s)& &(Hrs)&(MHz)&(mJy/beam) &(total \HI)&(velocity field)\\
&        &              &    &       &    &          &           &  \& PA (deg)  \\     
\hline
Arp004&01 48 25.7 -12 22 55 &1614 &Im&3.2&4.0 &1.50&20.0"$\times$20.0" &$-$\\
&        &              &    &       &    &          &           & $-$\\
      &                     &     &&3.2&4.0 &2.00& 46.7"$\times$ 38.9" & 46.7"$\times$38.9"\\
          &        &              &    &       &    &          &           & -29.1 \\
DDO015&01 49 40.2 -12 49 27 &1710 &Sm&3.2 &4.0&0.90 &20.0"$\times$20.0"&20.0"$\times$20.0"\\
&        &              &    &       &    &          &           & $-$\\
UGC3004 &04 17 19.0 +02 26 00 &3571 &S(?)&3.6 &8.0&1.00 &13.0"$\times$13.0"&13.0"$\times$13.0"\\
&        &              &    &       &    &          &           & $-$\\
UGC3005 &04 17 20.5 +02 27 01 &3215 &Scd &3.6&8.0&1.00 &13.0"$\times$13.0"&13.0"$\times$13.0"\\
&        &              &    &       &    &          &           &$-$ \\
UGC3006 &04 17 25.3 +02 22 16 &3664 &S0&3.6 &8.0&1.00 &$-$&$-$\\
&        &              &    &       &    &          &           &$-$ \\
UGC3014 &04 19 53.7 +02 05 36 &4214 &S(?)&6.0 &4.0&0.57 &15.0"$\times$15.0"&37.7"$\times$29.7"\\
&        &              &    &       &    &          &           & -36.4\\
MCG-03-34-41 &13 17 06.1 -16 15 08 &2651 &Sc&3.0 &4.0&1.20 &20.0"$\times$20.0"&20.0"$\times$20.0"\\
&        &              &    &       &    &          &           & $-$\\
MCG-03-34-04 &13 09 44.1 -16 36 08 &2619&S0 &3.3&8.0 &0.63 &25.0"$\times$25.0" &44.9"$\times$34.8"\\
&        &              &    &       &    &          &           & -12.3\\
NGC4997&13 09 51.7 -16 30 56 &2376 &S0&3.3 &8.0&0.63 &$-$&$-$\\
&        &              &    &       &    &          &           &$-$ \\
SGC1316.2-1722 &13 18 56.5 -17 38 06 &2495&Sm &2.6&4.0 &0.96 &20.0"$\times$20.0" &56.1"$\times$35.1"\\
&        &              &    &       &    &          &           & -15.1\\
SGC1317.2-1702 &13 19 54.8 -17 18 56 &2689&Sdm &3.0&4.0 &1.20&20.0"$\times$20.0" &20.0"$\times$20.0"\\
&        &              &    &       &    &          &           & $-$\\
NGC5031 &13 14 03.2 -16 07 23 &2839 &S0&3.3 &8.0&0.70 &$-$&$-$\\
&        &              &    &       &    &          &           & $-$\\
IC5269B &22 56 36.7 -36 14 59 &1667 &Scd&3.5 &4.0 &0.80 &20.0"$\times$20.0"&56.4"$\times$27.7"\\
&        &              &    &       &    &          &           & -19.6\\ 
\hline
\end{tabular}
\noindent M type : Morphological type, $\tau$ : Integration time, PA : Position Angle
\end{table}
}

\ctab{85}{tab1}{Expected surface matter densities for different morphological 
types (adapted from \citealt{H1Def} for using RC3 diameters instead of UGC diameters and taking h=1).
}{
\begin{center}
\begin{tabular}{p{4.3cm}r}
\hline
Morphological type ({\it M.T.})~ \& \hfill Index &log($\frac{M_{H_{I}}/D_{l}^{2}}{M_{\odot}/kpc^{2}}$)$_{pred} ~\pm$ s.d  \\ \hline 
 All \hfill 0 &6.89 $\pm$ 0.24 \\ 
 Elliptical,S0,S0/a \hfill 1 &6.69 $\pm$ 0.27 \\ 
Sa,Sab \hfill 2 &6.77 $\pm$ 0.32 \\ 
Sb \hfill 3 &6.91 $\pm$ 0.26 \\ 
Sbc \hfill 4 &6.93 $\pm$ 0.19 \\ 
Sc \hfill 5 &6.87 $\pm$ 0.19 \\ 
Scd,Sd,Irr,Sm,Sdm, dSp \hfill 6 &6.95 $\pm$ 0.17 \\ 
Pec \hfill 7 &7.14 $\pm$ 0.28 \\ \hline 
\end{tabular}
\end{center}
}



{\begin{table}
\caption{Observed and estimated parameters}
\begin{tabular}{lllllllr}
\hline
Galaxy  &angular    &HI mass &\HI deficiency &\dhbdt &Ram pressure&Evaporation &expected \HI~ \\ 
        &diameter(')&(\msun)    &       &      &\% stripped&in 1Gyr (\msun)    &deficiency\\       
\hline
Arp004 &2.8 &1.95$\times~ 10^{9}$ &0.15&$-$&34.4 &1.7$\times~ 10^{8}$ &0.23 \\
DDO015 &1.9 &1.80$\times~ 10^{8}$ &0.90&0.6&54.3 &7.7$\times~ 10^{7}$ &0.39\\
U3004 &1.2 &8.99$\times~ 10^{8}$ &0.40&0.9&1.6 &2.0$\times~ 10^{9}$ & 0.97\\
U3005 &1.2 &1.56$\times~ 10^{9}$ &0.12&1.6&$-$ &1.8$\times~ 10^{9}$ & $-$\\
U3014 &1.2 &2.34$\times~ 10^{9}$ &0.13&1.2&4.0 &2.3$\times~ 10^{9}$ & 0.64\\
MCG-03-34-41 &2.3 &1.66$\times~ 10^{9}$ &0.40&1.1&6.0 &1.4$\times~ 10^{9}$ &0.21 \\
MCG-03-34-04 &1.9 &8.00$\times~ 10^{9}$ &-0.64&1.8&$-$ &$-$ & $-$\\
SGC 1316.2-1722 &2.0 &1.07$\times~ 10^{9}$ &0.49&0.8&$-$ &1.1$\times~ 10^{9}$ & $-$\\
SGC 1317.2-1702 &1.9 &1.09$\times~ 10^{9}$ &0.51&0.8&$-$ &1.1$\times~ 10^{9}$ & $-$\\
IC5269B &4.1 &2.37$\times~ 10^{9}$ &0.42&1.1& 65.0 &4.6$\times~ 10^{8}$ & 0.56\\
\hline
\end{tabular}
\end{table}
}
{\begin{table}
\caption{GMRT Continuum observations}
\begin{tabular}{lllllllllr}
\hline
 Galaxy  &Total Flux Density    &rms & beam size& PA (deg)&Log of Radio&Log of FIR (60$\mu m$)&FIR (100$\mu m$)&q\\

           & from GMRT (mJy)  & (mJy/beam) & (")& &Luminosity&Luminosity&Flux (Jy)&\\
   &&& & & ($\rm {W Hz}^{-1}$) &(\lsun) &&\\         

\hline
UGC3004 &10.67 &0.09 & 4.0"$\times$4.0"&$-$ &21.46&9.51 &3.85 &2.37\\ 
UGC3014 &6.5 & 0.25  &13.1"$\times$10.7"&46.9 &21.39 &9.46 &2.00 &2.34\\
MCG-03-34-04 &21.79& 0.16  &6.2"$\times$2.6"&-11.2 &21.50&9.51 &5.21 &2.26\\
ARP004&$<$ 0.81 &0.27&4.0"$\times$4.0"&$-$ &$<$19.68 &8.48 &2.33 &3.19\\
MCG-03-34-41&$<$0.78&0.26&8.2"$\times$5.0"&-84.1 &$<$20.07 &8.83 &1.05 &3.01\\
\hline
\end{tabular}
\end{table}
}




\twocolumn 

\cfig{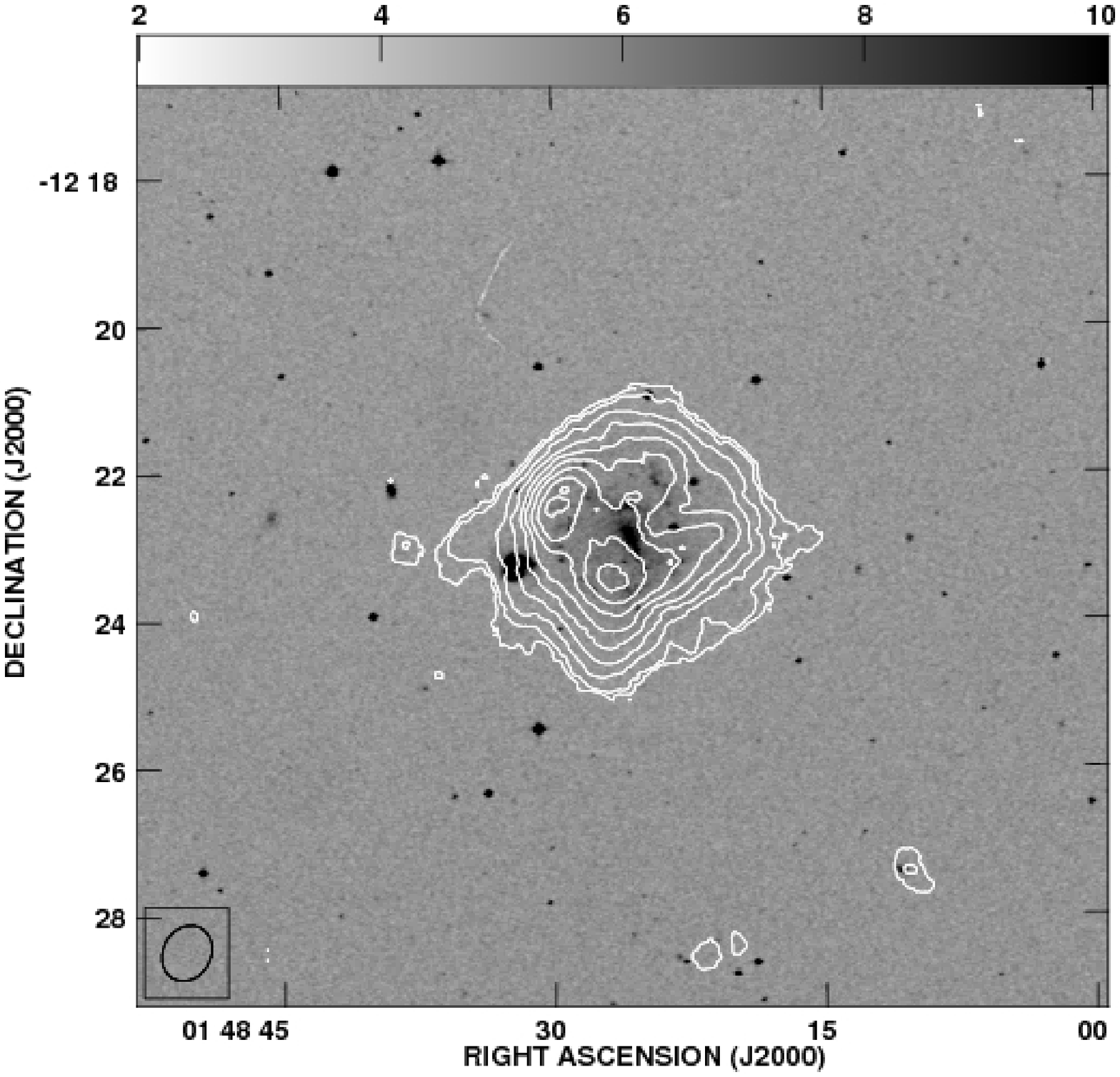}{Total \HI~ map (low resolution) : Arp004. The column density contours=1.8E+19/$cm^{2}$$\times$(3, 5, 10,15, 20, 25, 30, 35, 40)}{3.3}{}{-0.1}{0.0}{0.0}

\cfig{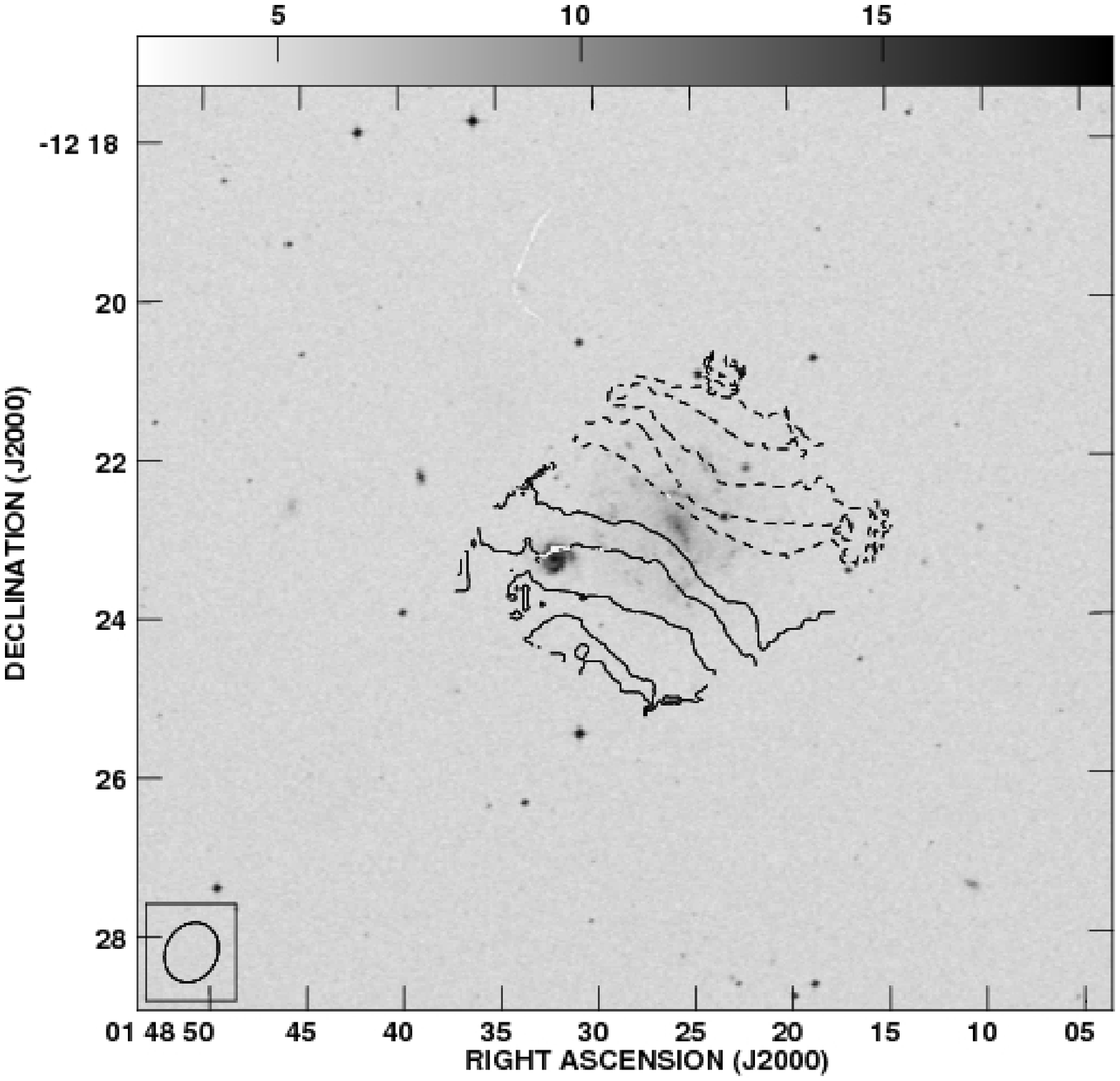}{\HI~ velocity field : Arp004. The velocity contours in km/s=1.0E+03$\times$(-50, -40, -30, -20, -10, 10, 20, 30, 40, 50)}{3.3}{}{-0.1}{0.0}{0.0}

\cfig{arphrpaped.ps}{Total \HI~ map (high resolution) : Arp004. The column density contours=3.3E+19/$cm^{2}$$\times$(3, 5, 7, 9, 11, 13, 15, 20 )}{3.9}{}{-0.4}{0.0}{0.0}

\cfig{ddo015paped.ps}{Total \HI~ map : DDO015. The column density contours=4.4E+19/$cm^{2}$$\times$(3, 5, 7, 9, 11, 13, 15)}{3.9}{}{-0.4}{0.0}{0.0}

\clearpage
\newpage

\cfig{ddoveled.ps}{\HI~ velocity field : DDO015. The velocity contours in km/s=1.0E+03$\times$(-40, -30, -20, -10, 10, 20, 30, 40)}{3.7}{}{-0.4}{0.0}{0.0}




\cfig{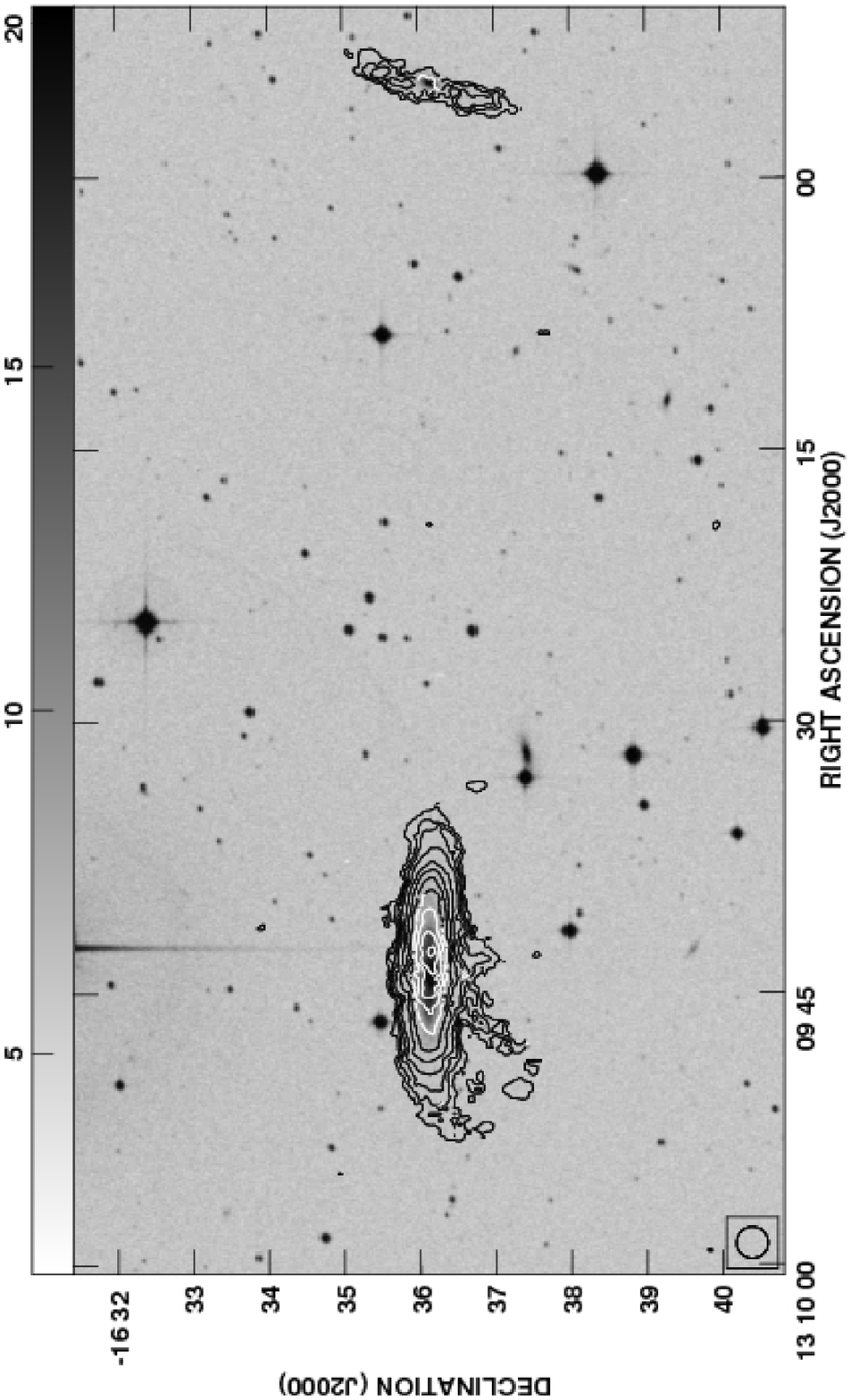}{Total \HI~ map : MCG-03-34-04. The column density contours=4.4E+19/$cm^{2}$$\times$(3, 5, 7, 10, 15, 20, 25, 30, 35, 40, 45, 50 )}{3.7}{}{-0.06}{-90.0}{0.0}

\cfig{mcg04veled.ps}{\HI~ velocity field : MCG-03-34-04 : The velocity contours in km/s=1.0E+03$\times$(-170, -160, -140, -120, -100, -80, -60, -40, -20, 20, 40, 60, 80, 100, 120, 130, 140)}{3.8}{}{-0.4}{-90.0}{0.0}

\cfig{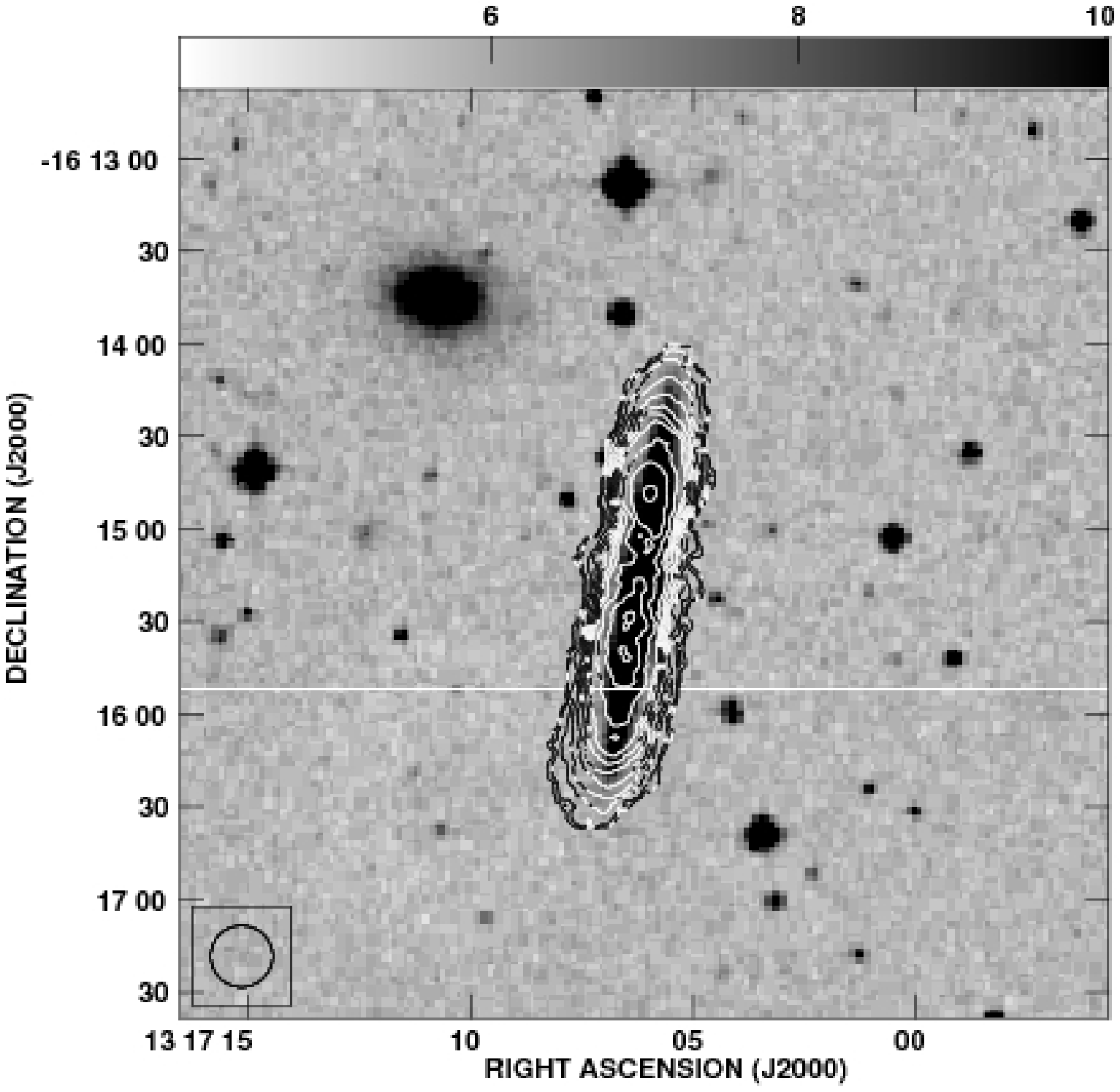}{Total \HI~ map : MCG-03-34-41 : \HI~ column density=4.1E+19/$cm^{2}$$\times$(3, 5, 10, 15, 20, 25, 30, 35, 45, 55, 65)  }{3.3}{}{-0.04}{0.0}{0.0}

\clearpage
\newpage

\cfig{mcg41veled.ps}{\HI~ velocity field : MCG-03-34-41 : \HI~ velocity contours in km/s=1.0E+03$\times$(-135, -125, -115, -95, -75, -55, -35, -15, -5, 5, 15, 35, 55, 75, 95, 115, 125, 135 )}{3.9}{}{-0.4}{0.0}{0.0}

\cfig{sgc02paped.ps}{Total \HI~ map : SGC1317.2-1702. The column density contours=4.4E+19/$cm^{2}$$\times$(3, 5, 7, 9, 11, 13, 15, 17, 19, 21, 23)}{3.9}{}{-0.4}{0.0}{0.0}

\cfig{sgc02veled.ps}{\HI~ velocity field : SGC1317.2-1702. The velocity contours in km/s=1.0E+03$\times$(-50, -40, -30, -20, -10, 10, 20, 30, 40, 50)}{3.9}{}{-0.4}{0.0}{0.0}

\cfig{sgc22paped.ps}{Total \HI~ map : SGC1316.2-1722. The column density contours=2.8E+19/$cm^{2}$$\times$(3, 5, 10, 15, 20, 25, 30 )}{3.9}{}{-0.4}{0.0}{0.0}

\clearpage
\newpage

\cfig{sgc22veled.ps}{\HI~ velocity field : SGC1316.2-1722. The velocity contours in km/s=1.0E+03$\times$(-35, -30, -25, -20, -15, -10, -5, 5, 10, 15, 20, 25, 30, 35, 40)}{4.0}{}{-0.4}{0.0}{0.0}

\cfig{u3014paped.ps}{Total \HI~ map : UGC3014. The column density contours=1.4E+20/$cm^{2}$$\times$(3, 5, 7, 10, 12, 15, 17)}{4.0}{}{-0.4}{0.0}{0.0}

\cfig{u3014veled.ps}{\HI~ velocity field : UGC3014. The velocity contours in km/s=1.0E+03$\times$(-110, -100, -80, -70, -60, -40, -20, -10, 10, 20, 40, 60, 70, 80, 100, 110)}{4.0}{}{-0.4}{0.0}{0.0}


\cfig{u3004veled.ps}{\HI~ velocity field : UGC3004. The velocity contours in km/s=1.0E+03$\times$(-70, -60, -50, -40, -30, -20, -10, 10, 20, 30, 40, 50, 60, 70 )}{4.0}{}{-0.4}{0.0}{0.0}

\clearpage
\newpage

\cfig{u3005veled.ps}{\HI~ velocity field : UGC3005. The velocity contours in km/s=1.0E+03$\times$(-100, -90, -80, -70, -60, -50, -40, -30, -20, -10, 10, 20, 30, 40, 50, 60, 70, 80, 90, 100 )}{4.0}{}{-0.4}{0.0}{0.0}

\cfig{u30045ed.ps}{Total \HI~ map : UGC3004 and UGC3005. The column density contours=9.1E+19/$cm^{2}$$\times$(3, 5, 7, 10, 15, 20, 25, 30, 35, 40, 45, 50, 55 )}{4.7}{}{-0.4}{0.0}{0.0}

\cfig{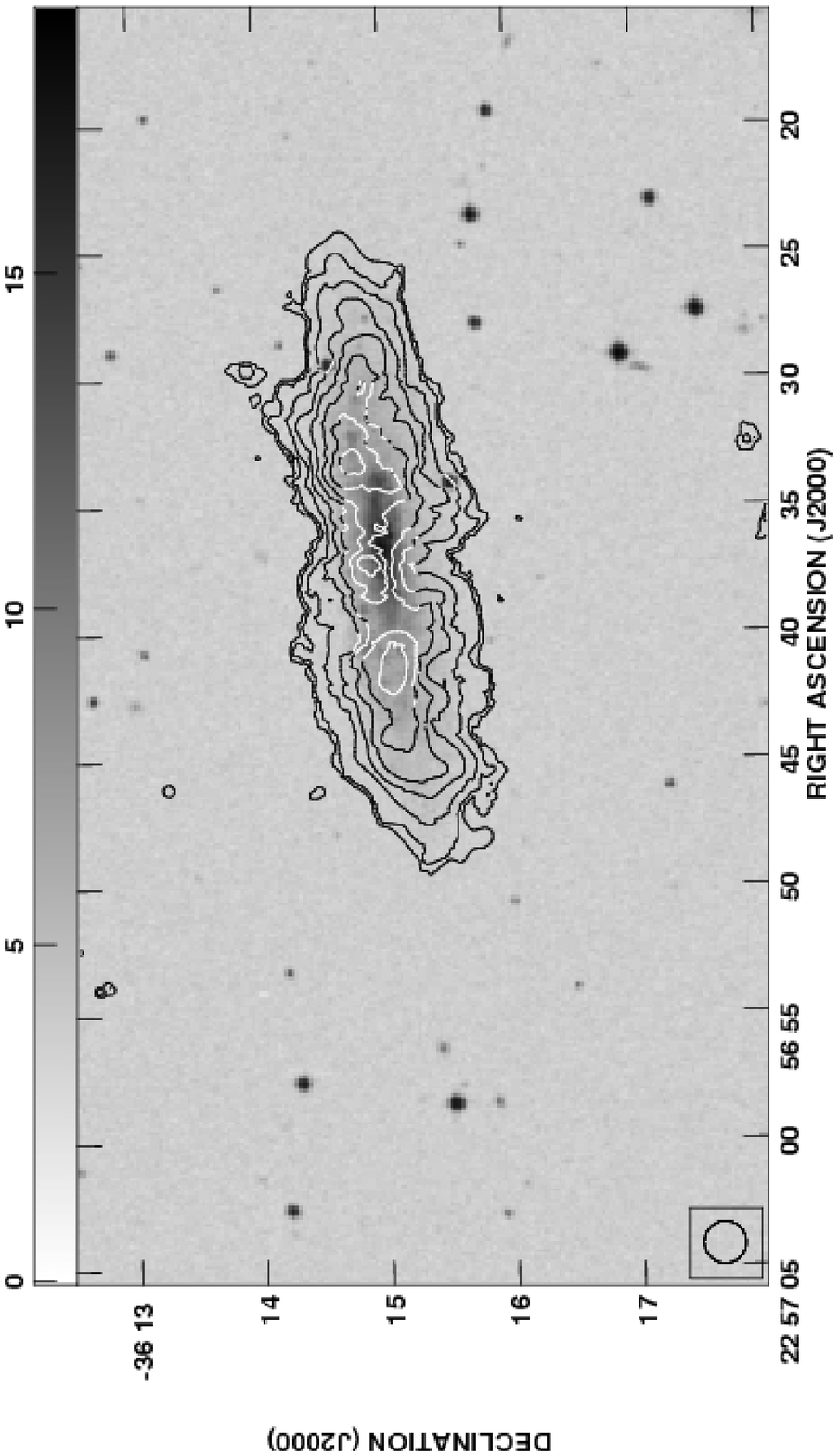}{Total \HI~ map : IC5269B. The column density contours=5.5E+19/$cm^{2}$$\times$(3, 5, 10, 15, 20, 25, 30, 35)}{4.2}{}{-0.06}{-90.0}{0.0}

\cfig{ic5269veled.ps}{\HI~ velocity field : IC5269B. The velocity contours in km/s=1.0E+03$\times$(-110, -100, -90, -80, -70, -50, -40, -30, -10, 10, 30, 40, 50, 70, 80, 90, 100, 110)}{3.9}{}{-0.4}{0.0}{0.0}

\clearpage
\newpage


\cfig{mcg04cont.ps}{20 cm radio continuum for MCG-03-34-04 contour levels = 1.500E-04 Jy$\times$(-5, -3, 3, 5, 7, 9, 13, 15)}{4.0}{}{-0.4}{0.0}{0.0}

\cfig{3004cont.ps}{20 cm radio continuum for UGC3004 contour levels = 8.000E-05 Jy $\times$ (-5, -4, -3, 3, 4, 5, 7, 9) }{4.0}{}{-0.4}{0.0}{0.0}
\cfig{3014cont.ps}{ 20 cm radio continuum for UGC3014 contour levels = 2.500E-04 JY $\times$ (-5, -3, 3, 4, 5, 6, 7)}{4.0}{}{-0.4}{0.0}{0.0}

\clearpage
\newpage
\onecolumn

\cfig{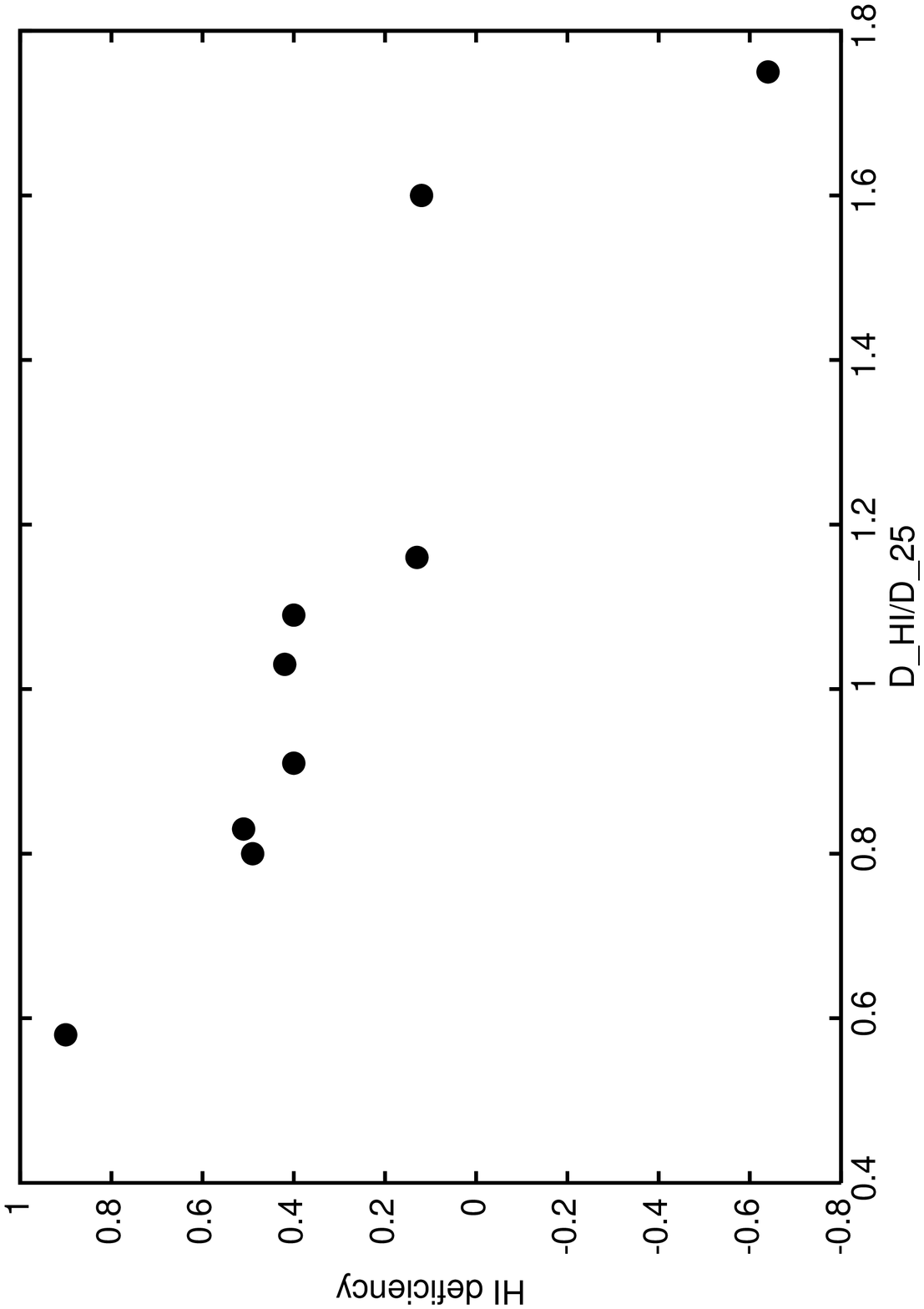}{\HI ~deficiency plotted against $D_{\HI}/D_{25}$ }{4.5}{}{-0.1}{-90.0}{0.0}

\cfig{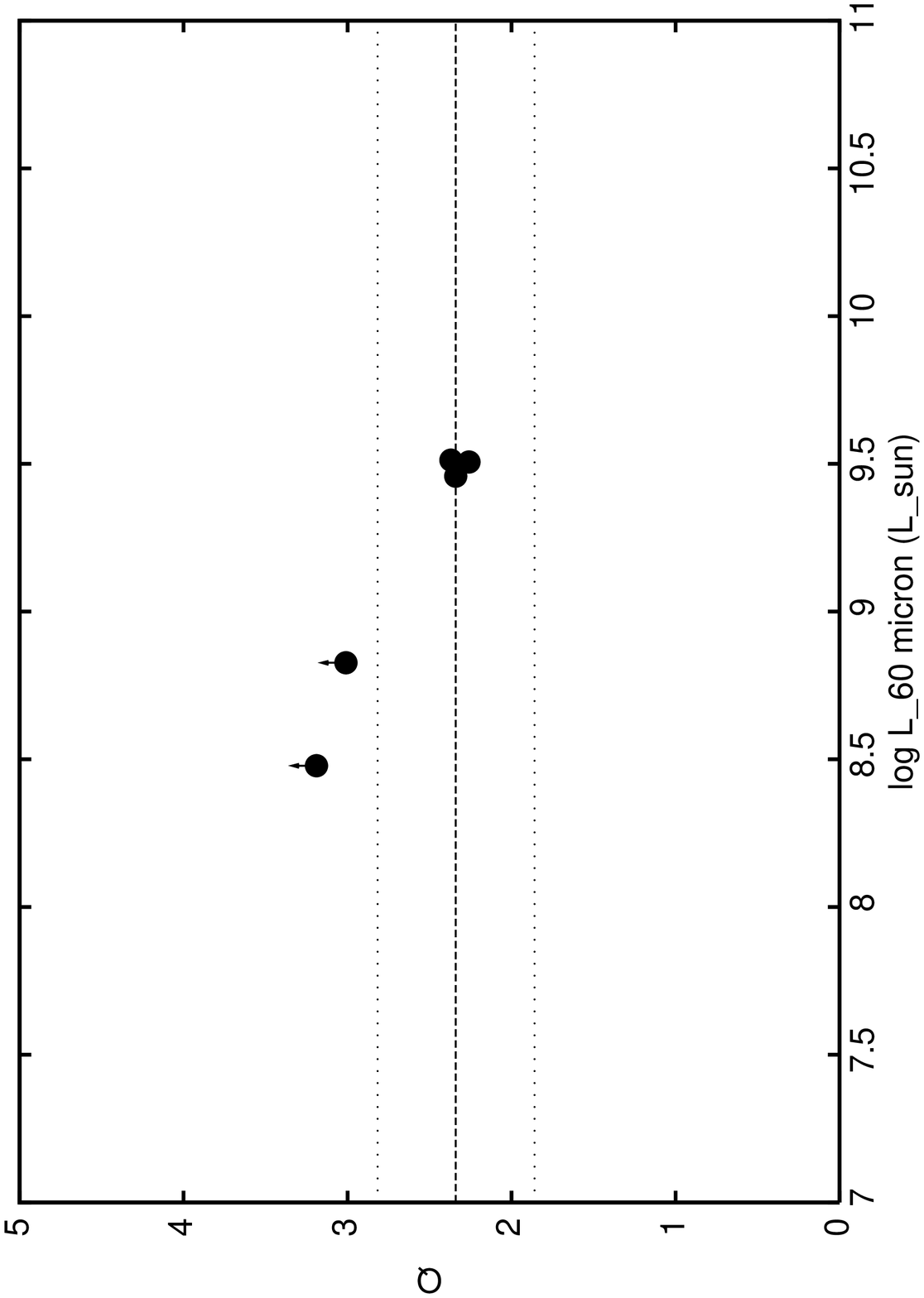}{The 'Q' value plotted against $\log \rm{L_{60\mu m}(L_{\odot})}$ }{4.5}{}{-0.05}{-90.0}{0.0}

\end{document}